\documentclass[prl,twocolumn,showpacs,preprintnumbers,amsmath,amssymb]{revtex4}

\usepackage{mathrsfs}%hanamoji
\usepackage{graphicx}% Include figure files
\usepackage{dcolumn}% Align table columns on decimal point
\usepackage{bm}% bold math
\usepackage{array}

\begin{document}

\title{Generalized Jarzynski Equality under Nonequilibrium Feedback Control}
\author{Takahiro Sagawa$^1$}
\author{Masahito Ueda$^{1,2}$}
\affiliation{$^1$Department of Physics, University of Tokyo,
7-3-1, Hongo, Bunkyo-ku, Tokyo, 113-8654, Japan \\
$^2$ERATO Macroscopic Quantum Control Project, JST, 2-11-16 Yayoi, Bunkyo-ku, Tokyo 113-8656, Japan
}
\date{\today}

\begin{abstract}
The Jarzynski equality is generalized to situations in which nonequilibrium systems are subject to a feedback control.  
The new terms that arise as a consequence of the feedback describe the mutual information content obtained by measurement and the efficacy of the feedback control.  Our results lead to a generalized fluctuation-dissipation theorem that reflects the readout information,  and can be experimentally tested using small thermodynamic systems.
We illustrate our general results by an introducing  ``information ratchet,''  which can transport a Brownian particle in one direction and  extract a positive work from the particle.
\end{abstract}


\maketitle

\textit{Introduction.} 
Since 1990's, nonequilibrium statistical mechanics has witnessed remarkable progress so that nonequilibrium dynamics can now be described in terms of  equalities~\cite{Equalities,Jarzynski}.  These equalities  have  been experimentally verified using small thermodynamic systems such as biomolecules or colloidal particles~\cite{Experiments}. 
A prominent example is  the Jarzynski equality~\cite{Jarzynski}:
\begin{equation}
\langle e^{- \beta (W - \Delta F)} \rangle = 1,
\label{Jarzynski}
\end{equation}
where $\langle W \rangle$ is the statistical average of work $W$ performed on a thermodynamic system,   $\Delta F$ is the free-energy difference, and $T \equiv (k_{\rm B}\beta)^{-1}$ is the temperature of the initial canonical distribution.  Equality~(\ref{Jarzynski}) is satisfied even when the final state of the system is far from equilibrium, i.e., even when we drive the system from $t = 0$ to $\tau$  by changing external  parameter $\lambda$ from $\lambda (0)$ to $\lambda (\tau )$ with a finite speed, where $\lambda$ describes, for example,  the volume of the gas or the frequency of an optical tweezer. 
The left-hand side of (\ref{Jarzynski}) involves all orders of cumulants of $W$.
In fact, the second law of thermodynamics ~\cite{Callen}
\begin{equation}
\langle W \rangle \geq \Delta F
\label{second-law}
\end{equation}
and the fluctuation-dissipation theorem result from the first and second cumulants of $W$, respectively~\cite{Jarzynski}.

Furthermore, recent advances in active control and precision measurement of small thermodynamic systems present new possibilities in nonequilibrium physics of small systems.
In particular, feedback control enhances our controllability of small thermodynamic systems~\cite{Ratchet,Feedback},
and plays a crucial role in biological and artificial nanomachines~\cite{Nanomachines}.
In addition to such engineering significances, feedback control on thermodynamic systems has also been a subject of active research in terms of the foundation of the second law of thermodynamics.
In fact, it is well understood that the role of the ``Maxwell's demon'' can be characterized as a feedback controller on thermodynamic systems~\cite{Maxwell,Sagawa-Ueda}.

Suppose that we perform a measurement on a stochastic thermodynamic system at time $t_{\rm m}$.  Let $\Gamma_{\rm m}$ be the phase-space point of the system at that time, $P[\Gamma_{\rm m}]$ its probability, and $y$ the measurement outcome.  
We assume that the measurement can involve a measurement error which is characterized by the conditional probability $P[y | \Gamma_{\rm m}]$ of obtaining outcome $y$ on condition that the state of the system is $\Gamma_{\rm m}$.  
For example, if a Gaussian noise is induced in the measurement, the conditional probability is given by $ P [y | \Gamma_{\rm m}] =( 2\pi N)^{-1/2} \exp ( - (y - \Gamma_{\rm m})^2 / 2N)$ with $N>0$ being the variance of the noise.  
The probability of obtaining outcome $y$ is given by  $P[y] = \int d\Gamma_{\rm m} P[y | \Gamma_{\rm m}]P[\Gamma_{\rm m}]$.
The information obtained by the measurement can be characterized by the mutual information~\cite{Cover-Thomas}, $\langle I \rangle \equiv \int d\Gamma_{\rm m}dy P[y | \Gamma_{\rm m}]P[\Gamma_{\rm m}] I[\Gamma_{\rm m}, y]$ with  $I [\Gamma_{\rm m}, y] \equiv \ln (P[y|\Gamma_{\rm m}] / P[y])$.
If we perform a feedback control, the control protocol of parameter $\lambda$ depends on the outcome $y$ after $t_{\rm m}$, which we write as $\lambda (t; y)$.
The introduction of the feedback control requires us to generalize the second law of thermodynamics~(\ref{second-law}) by including the mutual information $\langle I \rangle$ obtained by the feedback controller (or the ``demon'')~\cite{Sagawa-Ueda}:
\begin{equation}
\langle W \rangle \geq \Delta F - k_{\rm B}T \langle I \rangle.
\label{second-law-feedback} 
\end{equation}
Thus the work that needs to be performed on a thermodynamic system can be lowered by  feedback control.  

Now the crucial question is: Is it possible to generalize  the Jarzynski equality~(\ref{Jarzynski}) in the presence of feedback control such that we can find  more detailed information about nonequilibrium dynamics than inequality~(\ref{second-law-feedback}), as is the case for the original Jarzynski equality?  In this Letter we answer the question in the affirmative.

\textit{First Main Result.}
The generalized Jarzynski equality involves a term of information on the left-hand side:
\begin{equation}
\langle e^{-\beta (W-\Delta F) - I} \rangle =1,
\label{Jarzynski2}
\end{equation}
which will be proved later.  
We note that $\Delta F$ may depend on $y$ if $\lambda (\tau; y)$ does.
Our result  is applicable to classical stochastic processes that satisfy the local detailed balance (or the detailed fluctuation theorem)~\cite{Equalities}. 
Therefore, our result can be applied to a broad class of active control on small nonequilibrium systems.

The first cumulant of  Eq.~(\ref{Jarzynski2}) straightforwardly reproduce inequality~(\ref{second-law-feedback}) because of the concavity of the exponential function.  If all of the stochastic variables are Gaussian, the second cumulant leads to a generalized fluctuation-dissipation theorem including the term of the mutual information:
\begin{equation}
\langle \sigma + I \rangle  = \frac{1}{2}[\Delta (\sigma + I)]^2,
\label{fluctuation}
\end{equation}
where  $\sigma \equiv \beta (W - \Delta F)$ is the dissipation of work (or the entropy production), and $ [\Delta ( \sigma + I)]^2  \equiv \langle (\sigma + I)^2 \rangle - \langle \sigma + I \rangle^2$ is the variance of the sum of work and mutual information.  Therefore, the more information we get, the less dissipation the system will suffer.

\textit{Second Main Result.}
If we measure the left-hand side of the original Jarzynski equality~(\ref{Jarzynski}) in the presence of feedback control, the right-hand side is expected to differ from unity.  Let us write it as $\gamma$:
\begin{equation}
\langle e^{-\beta (W - \Delta F)} \rangle = \gamma.
\label{Jarzynski3}
\end{equation}
The crucial point is that we can directly measure $\gamma$ by using backward control protocols, and that $\gamma$ characterizes the efficacy of feedback control.  Thus the left-hand and right-hand sides of  Eq.~(\ref{Jarzynski3}) can be measured by the independent procedures.

We now discuss the properties of $\gamma$.  We  first note that the control protocol of $\lambda$ depends on measurement outcome $y$ at time $t > t_{\rm m}$ with feedback control.  In particular, if  the number of the possible outcomes is finite and given by $M$, we have $M$ kinds of protocols $\lambda (t; y)$ in the forward process.  Corresponding to each of them, we perform backward protocol $\lambda^\dagger (t; y) \equiv \lambda (\tau - t; y)$, which depends on $y$ only in $0 \leq t < \tau - t_{\rm m}$, by starting with the initial canonical distribution corresponding to parameter $\lambda^\dagger (0; y)$.
We stress that we do not perform any feedback control in the backward processes.  Instead, we drive the system depending on the forward outcome $y$ many times.
We then perform a measurement during the backward processes at time $\tau - t_{\rm m}$, and obtain outcome $y'$.
Let $P_{\lambda^\dagger (t; y)} [y']$ be the probability of obtaining outcome $y'$ with control protocol $\lambda^\dagger (t; y)$, which is normalized as  $\int dy' P_{\lambda^\dagger (t; y)} [y'] = 1$ for all $y$.
We then write the time-reversal $y$ as $y^\ast$;   if we only measure the momentum of the system, then $y_i^\ast = - y_i$; if we only measure the position of the system, then $y_i^\ast = y_i$.  
For a special case of $y' = y^\ast$, we use notation $P_{\lambda^\dagger(t; y)}[y^\ast]$, which is not necessarily unity.   Then  we can show that $\gamma$ is given by
\begin{equation}
\gamma = \int dy P_{\lambda^\dagger (t ; y)}[y^\ast].
\label{gamma}
\end{equation} 
As discussed in detail later,  to prove Eq.~(\ref{gamma}), we assume that the conditional probability satisfies  $P[y^\ast | \Gamma^\ast_{\rm m}] = P[y | \Gamma_{\rm m}]$.  
Here, $\Gamma^\ast_{\rm m}$ is the time-reversal of a phase-space point $\Gamma_{\rm m}$. For example,  if  $\Gamma_{\rm m} = (\bm r, \bm p)$ with position $\bm r$  and momentum $\bm p$, then $\Gamma^\ast_{\rm m} = (\bm r, - \bm p)$.  Physically,  $\gamma$ is the sum of the probabilities of obtaining the time-reversed outcomes with time-reversed protocols.  
Without feedback control, we have $\gamma = 1$ because $P_{\lambda^\dagger (t)}[ y^\ast]$ would then reduce to a single probability distribution.

The validity of Eq.~(\ref{Jarzynski3}) can be tested experimentally by measuring the left-hand side and the right-hand side independently;  we can measure $W$ and $\Delta F$ with forward processes, and determine $\gamma$ by performing time-reversed protocols $\lambda^\dagger (t;  y)$ many times for all possible outcomes $y$.    Once the validity of Eq.~(\ref{Jarzynski3}) has been confirmed, we can estimate the feedback efficacy $\gamma$ by only measuring $W$ and $\Delta F$ with forward protocols.  

We note that the effect of feedback control can be pronounced by Eq.~(\ref{Jarzynski3});   a small amount of work that satisfies $W < \Delta F$ makes an exponentially large amount of contribution on the left-hand side of Eq.~(\ref{Jarzynski3}).   In particular, with feedback control, the situation can occur in which  Eq.~(\ref{Jarzynski}) is violated while inequality~(\ref{second-law}) is still satisfied.  We will discuss such an example later. In such a situation, the feedback control only  affects the higher cumulants than the first order $\langle W \rangle$.

We next discuss the relationship between mutual information $I$ and parameter $\gamma$.  Let $C[X] \equiv \ln \langle e^{-X} \rangle$ be the cumulant generating function of a probability variable $X$.  From Eqs.~(\ref{Jarzynski2}) and (\ref{Jarzynski3}), and from an identity $\langle e^{-I} \rangle = 1$, we have $
C[\sigma + I] - C[\sigma] - C[I] = - \ln \gamma$.
The left-hand side of this equality characterizes the correlation between $\sigma$ and $I$, and therefore we find that $\gamma$ is a measure of the correlation between the dissipation and the information.  
In particular, if the joint distribution of $\sigma$ and $I$ is Gaussian,  we have  
\begin{equation}
\langle \Delta \sigma \Delta I \rangle = - \ln \gamma,
\label{correlation}
\end{equation}
where $\langle \Delta \sigma \Delta I \rangle \equiv \langle \sigma I \rangle - \langle \sigma \rangle \langle I \rangle$.
While $I$ only characterizes the information obtained by the measurement, $\gamma$ characterizes how efficiently we use the obtained information with feedback control. When $\gamma$ is large, we efficiently make dissipation $\sigma$ smaller by using the obtained information $I$, i.e., the more information $I$, the less dissipation $\sigma$.  We note that $I$ only depends on the measurement, but $\gamma$ depends both on the measurement and the feedback protocol. 

\textit{Examples.}
As an illustrative example, we consider  Eq.~(\ref{Jarzynski2}) for the Szilard engine~\cite{Maxwell}.  
The Szilard engine is a single-molecule ideal gas controlled by Maxwell's demon.
The gas is initially in thermodynamic equilibrium with a heat bath at temperature $T$. 
We  partition the  box into two boxes of equal volume. We then perform a measurement on the system to find out which box the molecule is in; the measurement outcome is ``left'' ($\equiv$`` L'') or ``right'' ($\equiv$ `` R''). By this measurement, we gain one bit ($= \ln 2$ nat) of information. 
When the outcome is ``R'',  we remove the left box and quasi-statically move the right one to the left.  
Finally, we expand the box to the right, and the state of the system S returns to the initial state. During the entire process, we extract $k_{\rm B}T \ln 2$ of work from the system with no free-energy change (i.e. $\Delta F = 0$).   
Since  $W = k_{\rm B}T \ln 2$ holds for all trajectories in the quasi-static limit and  $I = \ln 2$ holds for both ``L'' and ``R,''  we find that Eq.~(\ref{Jarzynski2})  holds for the case of the Szilard engine, that is, $\exp (-\beta \cdot (- k_{\rm B}T \ln 2)  -  \ln 2) =1$.

The backward process of the Szilard engine is described  as follows. The gas is initially in thermodynamic equilibrium, and we quasi-statically compress the box to the left.  The following step bifurcates into two branches depending on the  measurement outcome of the forward process.  If the outcome is ``L'', we do not move the box, and  perform the measurement of the position of the molecule.  Clearly, the outcome must be ``L'' with  unit probability: $P_{\lambda^\dagger (t; \rm L)} (\rm L) = 1$.  On the other hand, if the outcome is ``R'', we quasi-statically move the box to the right, and perform the measurement of the position of the molecule.  The outcome must be ``R'' with unit probability: $P_{\lambda^\dagger (t; \rm R)} (\rm R) = 1$. Finally, we remove the partition of the box and let the gas freely expand.   
We then obtain $\gamma = P_{\lambda^\dagger (t; \rm L)} (\rm L)  + P_{\lambda^\dagger (t; \rm R)} (\rm R) = 2$.
Therefore we find that Eq.~(\ref{Jarzynski3})  holds as $\exp (-\beta \cdot (- k_{\rm B}T \ln 2 - 0)) =2$.

We next discuss a model of ``information ratchet.''
We consider a one-dimensional Brownian particle in a harmonic potential.  Suppose that the particle is initially at thermal equilibrium in a potential $V_X(x) \equiv k(x-X)^2/2$, where $X$ is the center position of the potential.  We then measure the position $x$ of the particle and obtain outcome $y$. We assume that the measurement involves a Gaussian noise whose probability distribution is  $p(y-x) = (2\pi N)^{-1/2} \exp (- (y-x)^2 / 2N)$.  The joint probability of $x$ and $y$ with potential $V_X (x)$ is then given by
\begin{equation}
p_X(x,y) =\frac{1}{2\pi \sqrt{SN}} \exp \left( - \frac{(x-X)^2}{2S} - \frac{(y-x)^2}{2N} \right),
\end{equation}
where $S \equiv (k \beta)^{-1}$.
 Immediately after the measurement, we perform the following feedback control (see also FIG.~1 (a) for the case of $X=0$): if $y \geq X+L$ with $L>0$ being a constant, then we switch the potential to $V_{X+2L}(x)$; if $y<X+L$, we do nothing.  We next wait for relaxation of the particle.  When the probability distribution of the particle becomes a thermal equilibrium one,  we repeat the same feedback protocol by replacing $X$ by $X+2L$.  By performing this protocol many times, the average position of the particle moves to the right.  We note that $\Delta F = 0$ holds for this process. This one-way transportation of the particle looks like a ratchet model~\cite{Ratchet}.  However, the distinctive feature of the present model is that we do not need any asymmetry of the potential shape to drive the particle in one direction.   Moreover, we can even extract a  positive work during this transport if the measurement errors are small enough as discussed below.  Information obtained by  measurements enables the one-way transportation driven by feedback control so that we call this model ``information ratchet.''
\begin{figure}[htbp]
 \begin{center}
  \includegraphics[width=85mm]{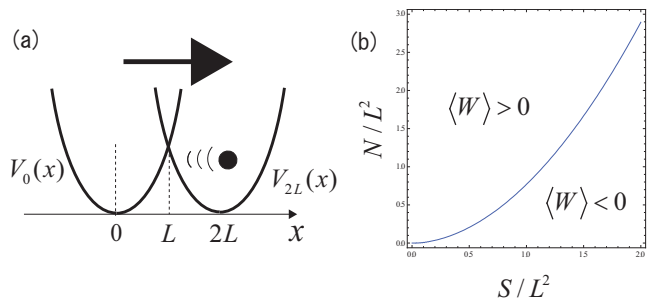}
 \end{center}
 \caption{(a) Schematic of the information ratchet with $X=0$, with which we transport a Brownian particle to the right solely by means of a feedback control.  (b) Regimes of $\langle W \rangle < 0$ and $\langle W \rangle >0$ on the $(S, N)$-plane.  The conventional second law~(\ref{second-law}) is violated only  in the regime of $\langle W \rangle <0$ with a large $S/N$-ratio,  while the original Jarzynski equality~(\ref{Jarzynski}) is not satisfied on the whole region.} 
\end{figure}

We discuss the energetics of the information ratchet for a single step.  The work performed on the particle at $x$ is given by $W(x) \equiv V_{X+2L} (x) - V_X (x)$ if $y\geq X+L$, and $0$ if $y<X+L$. Therefore the average work for each step is  
\begin{equation}
\begin{split}
\beta \langle W \rangle &=  \int_{-\infty}^\infty dx \int_{X+L}^\infty dy W (x) p_X(x,y) \\
&= \frac{L^2}{S} {\rm erfc} \left( \frac{L}{\sqrt{2(S+N)}} \right) - \frac{2L}{\sqrt{2\pi (S+N)}} e^{-L^2/ 2(S+N)},
\end{split}
\end{equation}
where ${\rm erfc}(t) \equiv (2/\sqrt{\pi})\int_t^\infty e^{-t'^2}dt'$.   We can show that $\langle W \rangle < 0$ holds for all $S$ if $N \to 0$, which means that we can extract a positive work during the one-way transportation if the measurement is error-free. 
Figure 1 (b) shows the regimes of $\langle W \rangle < 0$ and $\langle W \rangle >0$ on the $(S,N)$-plane.  
The mutual information is determined by the $S/N$-ratio: $\langle I \rangle = (1/2)\ln (1+S/N)$. Since $e^{-I} = p_X(x)p_X(y) / p_X(x,y)$ holds with $p_X(x) \equiv \int_{-\infty}^\infty dy p_X(x,y)$ and $p_X(y) \equiv \int_{-\infty}^\infty dx p_X(x,y)$, we obtain $\langle e^{-\beta W - I} \rangle = \int_{-\infty}^\infty  dx \int_{-\infty}^{X+L} dy 1\cdot p_X(x)p_X(y) +    \int_{-\infty}^\infty dx \int_{X+L}^\infty dy e^{-\beta W (x)} p_X(x)p_X(y) = 1$, and therefore Eq.~(\ref{Jarzynski2}) is satisfied  in this model.
We can also show that  $\langle e^{-\beta W} \rangle = \int_{-\infty}^\infty  dx \int_{-\infty}^{X+L} dy 1\cdot p_X(x,y) +    \int_{-\infty}^\infty dx \int_{X+L}^\infty dy e^{-\beta W (x)} p_X(x,y) = {\rm erfc} ( -L/\sqrt{2 (S+N)} )$, 
and
$\gamma =  \int_{-\infty}^{X+L} dy  p_X(y) +  \int_{X+L}^\infty dy  p_{X+2L}(y) = {\rm erfc} ( -L/\sqrt{2 (S+N)} )$. Therefore, Eq.~(\ref{Jarzynski3}) is also satisfied.  The efficacy parameter satisfies $\gamma > 1$ for all $(S,N)$ as long as $L>0$ so that the particle is transported to the right.  In addition,  $\gamma$ is a monotonically decreasing function of $N$, and  $\gamma \to 1$ holds with $N \to \infty$ which implies that the feedback control does not work at all because of an infinite amount of error.  We note that  $\langle W \rangle$ is positive for a regime of the small $S/N$-ratio even though $\gamma > 1$ always holds.  In this regime, the second law~(\ref{second-law}) is satisfied while the Jarzynski equality~(\ref{Jarzynski}) is violated as mentioned before.

\textit{Proof of the Main Results.}
  Let $\Gamma^\dagger (t) \equiv \Gamma^\ast (\tau -t)$ be the time-reversed trajectory of $\Gamma (t)$.  
With control protocol $\lambda (t; y)$ and $\lambda^\dagger (t; y)$, we denote the probability densities of trajectories $\Gamma (t)$ and $\Gamma^\dagger (t)$ as  $\mathcal P_{\lambda (t; y)}[\Gamma(t)]$ and  $\mathcal P_{\lambda^\dagger (t; y)}[\Gamma^\dagger(t)]$, respectively.  
They are normalized as $\int \mathcal P_{\lambda (t;  y)}[\Gamma(t)] \mathcal D [\Gamma (t)] = 1$ and   $\int \mathcal P_{\lambda^\dagger (t;  y)}[\Gamma^\dagger (t)] \mathcal D [\Gamma^\dagger (t)] = 1$, where  $\mathcal D [\Gamma (t)] = \mathcal D [\Gamma^\dagger (t)]$.
%We denote a probability density of trajectories  as $\mathcal P$  and that of phase-space points as $P$. 
It has been well-established that without any feedback control the local detailed balance holds for any control protocol~\cite{Equalities}, which is given by  $e^{-\sigma} = \mathcal P_{\lambda^\dagger (t; y)}[\Gamma^\dagger (t)] / \mathcal P_{\lambda (t; y)} [\Gamma (t)]$ with protocol $\lambda (t; y)$ with $y$ being fixed.

The joint distribution of $\Gamma (t)$ and $y$ is given by $P[y| \Gamma_{\rm m}] \mathcal P_{\lambda (t; y)} [\Gamma (t)]$.  Noting that $e^{-I} = P[y]/P[y|\Gamma_{\rm m}]$,  we have $\langle e^{- \sigma - I} \rangle = \int dy\mathcal D [\Gamma (t)] P[y | \Gamma_{\rm m}] \mathcal P_{\lambda (t; y)} [\Gamma (t)]  e^{-\sigma [\Gamma (t)]} P[y]/P[y| \Gamma_{\rm m}] = \int \mathcal D [\Gamma (t)] dy \mathcal P_{\lambda^\dagger (t; y)} [\Gamma^\dagger (t)] P[y] = 1$, which proves Eq.~(\ref{Jarzynski2}).

To prove Eq.~(\ref{gamma}), we use the assumption of  the time-reversal symmetry of the measurement, $P[y^\ast | \Gamma^\ast_{\rm m}] = P[y | \Gamma_{\rm m}]$.  The joint distribution of $\Gamma^\dagger (t)$ and $y'$ under the protocol $\lambda^\dagger (t; y)$ is given by $P[y' | \Gamma^\ast_{\rm m}] \mathcal P_{\lambda^\dagger (t; y)} [\Gamma^\dagger (t)]$ so that  $P_{\lambda (t; y)}[y'] = \int \mathcal D[\Gamma^\dagger (t)] P[y' | \Gamma^\ast_{\rm m}] \mathcal P_{\lambda^\dagger (t; y)} [\Gamma^\dagger (t)]$ for arbitrary $y$ and $y'$.  Therefore we obtain
\begin{equation}
\begin{split}
\langle e^{-\sigma} \rangle &= \int dy \mathcal D [\Gamma (t)] P[y | \Gamma_{\rm m}] \mathcal P_{\lambda (t; y)} [\Gamma (t)]  e^{-\sigma[\Gamma(t)]} \\
&= \int  dy\mathcal D [\Gamma^\dagger (t)] P[y^\ast | \Gamma^\ast_{\rm m}] \mathcal P_{\lambda^\dagger (t; y)} [\Gamma^\dagger (t)] \\
&= \int dy P_{\lambda^\dagger (t; y)}[y^\ast] 
\end{split}
\end{equation}
which proves Eq.~(\ref{gamma}).

In conclusion, we have generalized the Jarzynski equality to situations in which we perform a feedback control on a nonequilibrium dynamics.
The first generalization~(\ref{Jarzynski2}) includes the mutual information, and leads to the generalized second law~(\ref{second-law-feedback}) and the generalized fluctuation-dissipation theorem~(\ref{fluctuation}) corresponding to the first and second cumulants, respectively.
The second generalization~(\ref{Jarzynski3}) includes the efficacy parameter $\gamma$, which can be determined by backward processes, and characterizes the efficacy of feedback as shown in~(\ref{correlation}).
We have also illustrated the equalities by the Szilard engine and an information ratchet. 
We note that our results are consistent with the conventional  second law of thermodynamics because of the energy cost needed for the controller~\cite{Maxwell}.

\begin{acknowledgments}
This work was supported by a Grant-in-Aid for Scientific Research (Grant No.\ 17071005), and by a Global COE program ``Physical Science Frontier'' of MEXT, Japan. TS acknowledges JSPS Research Fellowships for Young Scientists (Grant No. 208038).
\end{acknowledgments}

\end{document}